\documentclass[submission,copyright,creativecommons]{eptcs}
\providecommand{\event}{SR2014} 
\usepackage{breakurl}           
\usepackage{comment}
\usepackage{amssymb,amsmath}
\usepackage{xspace}
\usepackage{tikz}
\usepackage{subfigure}
\usetikzlibrary{shapes,arrows,snakes,backgrounds,automata,patterns}

\title{Reasoning about Knowledge and Strategies:\\ Epistemic Strategy Logic}
\author{Francesco Belardinelli
\institute{Laboratoire IBISC -- Unversit\'e d'Evry}
\email{belardinelli@ibisc.fr}
}
\def\titlerunning{Epistemic Strategy Logic}
\def\authorrunning{F.~Belardinelli}

\newcommand{\aqis}{ECGM}

\newcommand{\free}{\textit{fr}}

\newtheorem{theorem}{Theorem}

\newtheorem{lemma}[theorem]{Lemma}
\newtheorem{definition}{Definition}

\newtheorem{claimAux}{Claim}

\newtheorem{exampleAux}{Example}

\long\def\eatpar#1{%
\ifx#1\par                      
\let\nextmove=\eatpar           
\else
\let\nextmove=#1
\fi
\nextmove
}

\def\qed{\hfill{\qedboxempty}      
  \ifdim\lastskip<\medskipamount \removelastskip\penalty55\medskip\fi}

\def\qedboxempty{\vbox{\hrule\hbox{\vrule\kern3pt
                 \vbox{\kern3pt\kern3pt}\kern3pt\vrule}\hrule}}

\def\qedfull{\hfill{\qedboxfull}   
  \ifdim\lastskip<\medskipamount \removelastskip\penalty55\medskip\fi}

\def\qedboxfull{\vrule height 4pt width 4pt depth 0pt}

\newcommand{{\incolumn}}[1]{\begin{tabular}[c]{c} #1 \end{tabular}}
\newcommand{{\incolumnmath}}[1]{\begin{array}[c]{c} #1 \end{array}}

\begingroup
\catcode`\~=11
\gdef\urltilde{\lower 0.6ex\hbox{~}}
\endgroup

\newcommand{\A}{\mathcal{A}} 
 
 \newcommand{\F}{\mathcal{F}}

 \renewcommand{\L}{\mathcal{L}}
 \newcommand{\N}{\mathcal{N}}
 \renewcommand{\P}{\mathcal{P}}
\newcommand{\Q}{\mathcal{Q}} 
 \newcommand{\T}{\mathcal{T}}

\newcommand{\defterm}[1]{\mbox{\underline{\it\smash{#1}\vphantom{\lower.1ex\hbox
{x}}}}}

\newcommand{\set}[1]{\{#1\}}                      

\newcommand{\tup}[1]{\langle #1\rangle}            

\newcommand{\myi}{\emph{(i)}\xspace}
\newcommand{\myii}{\emph{(ii)}\xspace}
\newcommand{\myiii}{\emph{(iii)}\xspace}
\newcommand{\myiv}{\emph{(iv)}\xspace}
\newcommand{\myv}{\emph{(v)}\xspace}
\newcommand{\myvi}{\emph{(vi)}\xspace}

\newcommand{\Nat}{{\rm I\kern-.23em N}}

\newcommand{\commentout}[1]{}

\usepackage{marginnote}
\edef\marginnotetextwidth{\the\textwidth}

\begin{document}
\maketitle


\begin{abstract}
In this paper we introduce Epistemic Strategy Logic (ESL), an
extension of Strategy Logic with modal operators for individual
knowledge. This enhanced framework allows us to represent explicitly
and to reason about the knowledge agents have of their own and other
agents' strategies.  We provide a semantics to ESL in terms of
epistemic concurrent game models, and consider the corresponding
model checking problem.  We show that the complexity of model checking
ESL 
is not worse than
(non-epistemic) Strategy Logic.
\end{abstract}


\section{Introduction}


Formal languages to represent and reason about strategies and
coalitions are a thriving area of research in Artificial Intelligence
and multi-agent system \cite{BullingDJ10,GorankoJ04,Pauly02}.
Recently, a wealth of multi-modal logics have appeared, which allow to
formalise complex strategic abilities and behaviours of individual
agents and groups \cite{atl02,ChatterjeeHP10}.
In parallel to these developments, in knowledge representation there
is a well-established tradition of extending logics for reactive
systems with epistemic operators to reason about the knowledge agents
have of systems evolution.  These investigations began in the '80s
with contributions on combinations of linear- and branching-time
temporal logics with multi-agent epistemic
languages \cite{HalpernV86,HalpernV89,fhmv:rak}. 
Along this line of research, \cite{HoekW03a} introduced
alternating-time temporal epistemic logic (ATEL), an extension of ATL
with modalities for individual knowledge.
The various flavours of logics of time and knowledge have been
successfully applied to the specification of distributed and multi-agent
systems in domains as diverse as security protocols, UAVs, web
services, and e-commerce, as well as to verification by model
checking \cite{GammieM04,LomuscioQR09}.

In this paper we take inspiration from the works above and pursue
further this line of research by introducing Epistemic Strategy Logic,
an extension of Strategy Logic (SL) \cite{ChatterjeeHP10,MogaveroMV11}
that allows agents to reason about their strategic abilities.  The extension here proposed is naive in the sense that it suffers many of the shortcomings of its
relative ATEL \cite{Jamroga04}. Nonetheless, we reckon that it constitutes an
excellent starting point to analyse the interaction of knowledge and
strategic abilities in a language, such as SL,  that explicitly allow for quantification on
strategies.



\textbf{Related Work.} 
This paper builds on previous contributions on Strategy Logic.  SL has
been introduced in \cite{ChatterjeeHP10} for two-player concurrent
game structures (CGS). In \cite{MogaveroMV11} the semantics has been
extended to a multi-player setting. Also, \cite{MogaveroMV11}
introduced bind operators for strategies in the syntax.  In
the present contribution we consider multi-agent CGS in line with
\cite{MogaveroMV11}. However, we adopt an agent-based perspective and consider agents with possibly different actions and protocols \cite{fhmv:rak}. 
Also, our language do not include bind operators 
to avoid the formal machinery associated with these operators.
We leave such an extension for future and more comprehensive work.
Finally, the model checking results in Section~\ref{MCesl} are
inspired by and use techniques from \cite{MogaveroMV11}.

Even though to our knowledge no epistemic extension of SL has been
proposed yet, the interaction between knowledge and strategic
reasoning has been studied extensively, especially in the context of
alternating-time temporal logic. 
An extension of ATL with knowledge operators, called ATEL, was
put forward in \cite{HoekW03a}, and immediately imperfect
information variants of this logic were considered
in \cite{JamrogaH04}, which introduces alternating-time temporal observational
logic (ATOL) and ATEL-R*, as well as uniform strategies.
Notice that \cite{JamrogaH04} also analyses the distinction between
{\em de re} and {\em de dicto} knowledge of strategies; this
distinction will also be considered later on in the context of
Epistemic Strategy Logic.
Further,
\cite{JamrogaA07} enriches ATL with a
constructive notion of knowledge.
As regards (non-epistemic) ATL, more elaborate notions of strategy have been considered. In \cite{AgotnesGJ07} commitment in strategies has been analysed; while \cite{Jonker03} introduced a notion of ``feasible'' strategy.
In future work it might be worth exploring to what extent the
theoretical results available for the various flavours of ATEL
transfer to ESL.

\textbf{Scheme of the paper.} In Section~\ref{preliminaries}  we introduce the epistemic concurrent game models (\aqis), which are used in Section~\ref{esl} to provide a semantics to Epistemic Strategy Logic (ESL).
In Section~\ref{MCesl} we consider the model checking problem for this
setting and state the corresponding complexity results.
Finally, in Section~\ref{conc} we discuss the results and point to
future research.
For reasons of space, all
proofs are omitted.  An extended version of this paper with complete
proofs is available \cite{Belardinelli14tech}.


\section{Epistemic Concurrent Game Models} 



\label{preliminaries}

In this section we present the epistemic concurrent game models
(\aqis), an extension of concurrent game structures \cite{atl02,HoekW03a}, 
 starting with the notion of {\em agent}.
\begin{definition}[Agent]\label{def:ags}
An {\em agent} is a tuple
$i = \tup{L_i, Act_i, Pr_i}$ such that
\myi $L_i$ is the set of {\em local states} $l_i, l'_i, \ldots$;
\myii $Act_i$ is the finite set of \emph{actions} $\sigma_i, \sigma'_i, \ldots$; and
\myiii $Pr_i: L_i \mapsto 2^{Act_i}$ is the {\em protocol function}.
\end{definition}

Intuitively, 
each agent $i$ is
situated in some local state $l_i \in L_i$, representing her local
information, and performs the actions in $Act_i$ according to the
protocol function $Pr_i$ \cite{fhmv:rak}.  Differently from
\cite{MogaveroMV11}, we assume that agents have possibly different 
actions and protocols.
To formally describe the interactions between agents, we introduce their
synchronous composition.  Given a set $AP$ of atomic propositions and
a set $Ag=\set{i_0, \ldots, i_n}$ of agents,
we define the set $L$ of global
states $s, s', \ldots$ (resp.~the set $Act$ of joint actions $\sigma,
\sigma', \ldots$) as the cartesian product $L_0 \times \ldots \times L_n$
(resp.~$Act_0 \times \ldots \times Act_n$).  In what follows we denote the $j$th
component of a tuple $t$ as $t_j$ or, equivalently, as $t(j)$.
\begin{definition}[\aqis]\label{def:sys-db-ag}
Given a set $Ag=\set{i_0, \ldots, i_n}$ of agents $i =\tup{L_i, Act_i,
  Pr_i}$,
an {\em epistemic
  concurrent game model} is a tuple $\P = \tup{Ag, s_0, \tau, \pi}$
such that
\myi $s_0 \in L$ is the {\em
  initial global state};
\myii $\tau: L \times Act \mapsto
  L$ is the {\em global transition function}, where $\tau(s,\sigma)$
  is defined iff $\sigma_i \in Pr_i(l_i)$ for every $i \in Ag$;
and 
\myiii $\pi : AP \mapsto 2^{L}$ is the {\em
  interpretation function} for atomic propositions in $AP$.
\end{definition}

The transition function $\tau$ describes the evolution of the
\aqis\ from the initial state $s_0$.
We now introduce some 
notation that will be used in the rest of the paper.  The {\em
transition relation} $\to$ on global states is defined as $s
\to s'$ iff there exists $\sigma \in Act$ 
s.t.~$\tau(s, \sigma) = s'$.  
A \emph{run} $\lambda$ from a
state $s$, or $s$-run, is an infinite sequence $s^0 \to s^1 \to \ldots $,
where $s^0=s$.  
For $n, m \in \mathbb{N}$, with $n \leq m$, we define
$\lambda(n) = s^n$ and $\lambda[n,m] = s^n,s^{n+1}, \ldots, s^{m}$.  A
state $s'$ is \emph{reachable from $s$} if there exists an $s$-run
$\lambda$
s.t.~$\lambda(i) = s'$ for some
$i\geq 0$.
We define 
$S$ as the set  of states reachable from the initial state $s_0$.
Further, let $\sharp$ be a placeholder for arbitrary individual actions. Given a
subset $A \subseteq Ag$ of agents, an {\em $A$-action} $\sigma_A$ is
an $|Ag|$-tuple s.t.~\myi $\sigma_A(i) \in Act_i$ for $i \in A$,
and \myii $\sigma_A(j) = \sharp$ for $j \notin A$.  
Then, $Act_A$ is the set of all $A$-actions and $D_A(s)
= \{ \sigma_A \in Act_A \mid$ for every $i \in A, \sigma_i \in
Pr_i(l_i) \}$ is the set of all $A$-actions enabled at $s
=\tup{l_0,\ldots,l_n}$.
A joint action $\sigma$ {\em extends} an
$A$-action $\sigma_A$, or $\sigma_A
\sqsubseteq \sigma$, iff $\sigma_A(i) = \sigma(i)$ for all $i \in A$.
The {\em outcome $out(s, \sigma_{A})$ of action $\sigma_{A}$ at state $s$} is the
set of all states $s'$ s.t.~there exists a joint action $\sigma
 \sqsupseteq \sigma_A$ and $\tau(s, \sigma) =
s'$.
%
Finally, two global states
$s=\tup{l_0,\ldots,l_n}$ and $s'=\tup{l'_0,\ldots,l'_n}$ are
\emph{indistinguishable} for agent $i$, or $s
\sim_i s'$, iff $l_i = l'_i$ \cite{fhmv:rak}. 


\section{Epistemic Strategy Logic} \label{esl}


We now introduce Epistemic Strategy Logic as a specification
language for \aqis.  Hereafter we consider a set $Var_i$ of strategy variables
$x_i, x'_i, \ldots$, for every agent $i \in Ag$. 
\begin{definition}[ESL] \label{def:FO-CTLK}
For $p \in AP$, $i \in Ag$ and $x_i \in Var_i$,
 the ESL formulas $\phi$ are defined in BNF as follows: 
\begin{eqnarray*}
\phi   & ::= &  
p  \mid \neg \phi \mid \phi \to \phi \mid X \phi \mid \phi
  U \phi \mid  K_i \phi 
\mid \exists x_i \phi
\end{eqnarray*}
\end{definition}

The language ESL is an extension of the Strategy
Logic in \cite{ChatterjeeHP10} to a multi-agent setting, 
including an epistemic operator $K_i$ for each $i \in Ag$.
Alternatively, ESL can be seen as the epistemic extension of the Strategy
Logic in \cite{MogaveroMV11}, minus the bind operator. We do not consider
 bind operators in ESL for ease of presentation. 
%
The ESL formula 
$\exists x_i
\phi$ is read as 
``agent $i$ has some strategy to achieve $\phi$''. The interpretation
of LTL operators $X$ and $U$ is standard.
%
The epistemic formula $K_i \phi$ 
intuitively means that ``agent $i$ knows $\phi$''.  The other
propositional connectives and LTL operators, as well as the strategy
operator $\forall$, can be defined as standard.
%
%
Also, 
notice that 
we can introduce the {\em nested-goal} fragment ESL[NG],
the {\em boolean-goal} fragment ESL[BG], and the {\em one-goal}
fragment ESL[1G] in analogy to SL \cite{MogaveroMV11}. 
Further, the {\em free} variables $\free(\phi) \subseteq Ag$ of an ESL
formula $\phi$ are inductively defined as follows:
\begin{tabbing}
 $\free(\forall x_i \phi) = \free(\exists x_i \phi)$ \ \ \= $=$ \ \ \= $\free(\phi) \setminus \{ i \}$ \kill 
 $\free(p)$  \> $=$ \> $\emptyset$\\
 $\free(\neg \phi) = \free(K_i \phi)$ \> $=$ \> $\free(\phi)$\\  
 $\free(\phi \to \phi')$ \> $=$ \> $\free(\phi) \cup \free(\phi')$\\ 
 $\free(X \phi) = \free(\phi U \phi')$ \> $=$ \> $Ag$\\  
$\free(\exists x_i \phi)$ \> $=$ \> $\free(\phi) \setminus \{ i \}$
\end{tabbing}
A {\em sentence} is a formula $\phi$ with $\free(\phi)
= \emptyset$, and the set $bnd(\phi)$ of bound variables  is defined as
$Ag \setminus \free(\phi)$.

To provide a semantics to ESL formulas in terms of \aqis, we introduce the
notion of strategy.
\begin{definition}[Strategy] \label{strategy}
Let $\gamma$ be an ordinal s.t.~$1 \leq \gamma \leq \omega$ and
$A \subseteq Ag$ a set of agents. A {\em $\gamma$-recall
$A$-strategy} is a function $F_A[\gamma] : \bigcup_{1 \leq n <
1+\gamma} S^n \mapsto \bigcup_{s \in S} D_A(s)$
s.t.~$F_A[\gamma](\kappa) \in D_A(last(\kappa))$ for every $\kappa \in
\bigcup_{1 \leq n < 1+\gamma} S^n$, where $1+\gamma = \gamma$ for $\gamma = \omega$ and $last(\kappa)$ is the last
element of $\kappa$.
\end{definition}

Hence, a $\gamma$-recall $A$-strategy returns an enabled $A$-action
for every sequence $\kappa \in
\bigcup_{1 \leq n < 1+\gamma} S^n$ of states of length at most $\gamma$.
Notice that for $A = \{ i\}$, $F_A[\gamma]$ can be seen as a function
from $\bigcup_{1 \leq n < 1+\gamma} S^n$ to $Act_i$
s.t.~$F_A[\gamma](\kappa) \in Pr_i(last(\kappa))$ for
$\kappa \in
\bigcup_{1 \leq n < 1+\gamma} S^n$.
In what follows we write $F_i[\gamma]$ for $F_{\{ i \}}[\gamma]$. 
Then, for $A = \{i_0, \ldots, i_m \} \subseteq Ag$, $F_A[\gamma]$ is equal to 
$F_{i_0}[\gamma] \times \ldots \times F_{i_m}[\gamma]$, where for every
$\kappa \in
\bigcup_{1 \leq n < 1+\gamma} S^n$,
$(F_{i_0}[\gamma] \times \ldots \times F_{i_m}[\gamma])(\kappa)$ is
defined as the set of actions $\sigma \in \bigcup_{s \in S} D_A(s)$
s.t.~$\sigma_i = F_i[\gamma](\kappa)$ if $i \in A$, $\sigma_i
= \sharp$ otherwise.  Therefore, a group strategy is the composition
of its members' strategies.
%
Further, the {\em outcome} of strategy $F_A[\gamma]$ at state $s$,
or $out(s, F_A[\gamma])$, is the set of all $s$-runs $\lambda$
s.t.~$\lambda(i + 1) \in out(\lambda(i), F [\gamma](\lambda[j, i]))$
for all $i \geq 0$ and $j = max(i -\gamma + 1, 0)$.
Depending on $\gamma$ we can define positional strategies, strategies
with perfect recall, etc.~\cite{GorankoJ04}. However, these different
choices do not affect the following results, so we assume that
$\gamma$ is fixed and omit it.  Moreover, by Def.~\ref{strategy}
it is apparent that agents have perfect information, as their strategies
are determined by global states \cite{BullingDJ10}; we leave
contexts of imperfect information for future research.

Now let $\chi$ be an assignment that maps 
each agent $i \in Ag$ to an $i$-strategy $F_i$.
For $Ag = \{i_0, \ldots, i_n \}$, we denote
$\chi(i_0) \times \ldots \times \chi(i_n)$ as $F^{\chi}$, that is,
the $Ag$-strategy s.t.~for every $\kappa \in
\bigcup_{1 \leq n < 1+\gamma} S^n$, $F^{\chi}(\kappa) = \sigma \in \bigcup_{s \in
S} D_{Ag}(s)$ iff $\sigma_i = \chi(i)(\kappa)$ for every $i \in Ag$.
Since $|out(s, F^{\chi})| = 1$, we simply write $\lambda  = out(s, F^{\chi})$.
Also, 
$\chi^{i}_{F_i}$ denotes the assignment s.t.~\myi for all agents
$j$ different from $i$, $\chi^{i}_{F_i}(j) = \chi(j)$, and \myii
$\chi^{i}_{F_i}(i) = F_i$.  
\begin{definition}[Semantics of ESL] \label{def:fo-ctl-sem}
We define whether an \aqis\ $\P$ {\em satisfies} a formula $\varphi$ 
at state $s$ according to assignment $\chi$,
or $(\P,s, \chi)\models \varphi$, as follows (clauses for propositional connectives are straightforward and thus omitted): 
{\small
\begin{tabbing}
 $ (\P,s, \chi)\models p$ \ \ \ \ \ \ \ \ \ \ \ \ \ \= iff \ \ \ \ \ \ \= $s \in \pi(p)$\\
 $(\P,s,\chi) \models X \psi$ \>  iff \>  for 
 $\lambda = out(s, F^{\chi})$, $(\P, \lambda(1),\chi)
  \models \psi$\\
 $(\P,s,\chi) \models \psi U \psi'$ \>  iff \>  
for  $\lambda = out(s, F^{\chi})$  there is $k\geq 0$ s.t.~$(\P,\lambda(k),\chi) \models\psi'$\\
\> \> and 
$0\leq j< k$ implies
$(\P,\lambda(j),\chi) \models \psi$\\
 $(\P,s,\chi) \models K_i \psi$ \>  iff \>  for all 
 $s \in S$, $s \sim_i s'$ implies $(\P, s',\chi) \models \psi$\\
 $(\P,s,\chi) \models \exists x_i \psi$  \> iff \>
 there exists an $i$-strategy $F_i$ s.t.~$(\P,s,\chi^{i}_{F_i}) \models \psi$
\end{tabbing}
}
\end{definition}
An ESL formula $\varphi$ is {\em satisfied} at state $s$,
or $(\P, s) \models
\varphi$, if $(\P, s, \chi) \models \varphi$ for all assignments $\chi$;
$\varphi$ is {\em true} in $\P$, or $\P \models \varphi$, if $(\P,
s_0)
\models \varphi$. 
The satisfaction of formulas is independent from bound variables, that
is, $\chi(\free(\phi)) = \chi'(\free(\phi))$ implies that $(\P,
s, \chi) \models \phi$ iff $(\P, s, \chi') \models \phi$. In
particular, the satisfaction of sentences is independent from assignments.

We can now state the model checking problem for ESL.
\begin{definition}[Model Checking Problem]
        Given an \aqis\ $\P$ and an ESL formula $\phi$, determine
        whether there exists an assignment $\chi$ s.t.~$(\P,
        s_0, \chi) \models \phi$.
\end{definition}

Notice that, if $y_1, \ldots, y_m$ is an enumeration of $\free(\phi)$, 
then the model checking problem
amounts to check whether $\P \models \exists y_1, \ldots, \exists y_m \phi$, where $\exists y_1, \ldots, \exists y_m \phi$ is a sentence.

Hereafter we illustrate the formal machinery introduced thus far
with a toy example.\\

\begin{figure} 
	\centering
\begin{tikzpicture}[auto,node distance=2.3cm,->,>=stealth',shorten
>=1pt,semithick] \label{fig1.1}
{\small

\tikzstyle{every state}=[fill=white,draw=black,text=black,minimum
  size=0.4cm, rectangle, rounded corners = 0.2 cm]

    \node[state, label = above: $s_{0}$] (ee) at(0,0) {$(\epsilon_A,\epsilon_B)$};
    \node[state, label = left: $s_{0\lambda}$] (0e) at(-2.5,-1) {$(0,\lambda)$};
    \node[state, label = right: $s_{1\lambda}$] (1e) at(2.5,-1) {$(1,\lambda)$};
    \node[state, label = left: $s_{00}$] (00) at(-4,-2.4) {$(0,0)$};
    \node[state, label = right: $s_{01}$, node distance = 1.8 cm] (01) at(-2.5,-2.4) {$(0,1)$};
    \node[state, label = left: $s_{10}$, node distance = 1.8 cm] (10) at(2.5,-2.4) {$(1,0)$};
    \node[state, label = right: $s_{11}$] (11) at(4,-2.4) {$(1,1)$};
    \path (ee) edge node[above left]  {$(set(0), wait)$} (0e)
    (ee) edge node[above right]  {$(set(1), wait)$} (1e)
    (0e) edge node[left]  {$(skip, set(0))$} (00)
    (0e) edge node[right]  {$(skip, set(1))$} (01)
    (1e) edge node[left]  {$(skip, set(0))$} (10)
    (1e) edge node[right]  {$(skip, set(1))$} (11)
    (0e) edge[dashed,-] node[above]  {$B$} (1e);
}
\end{tikzpicture}
\caption{the \aqis\ $\Q$. Transitions from $s_{00}$, $s_{01}$, $s_{10}$, and $s_{11}$ to $s_{0}$ are omitted.} \label{fig1}
\end{figure}
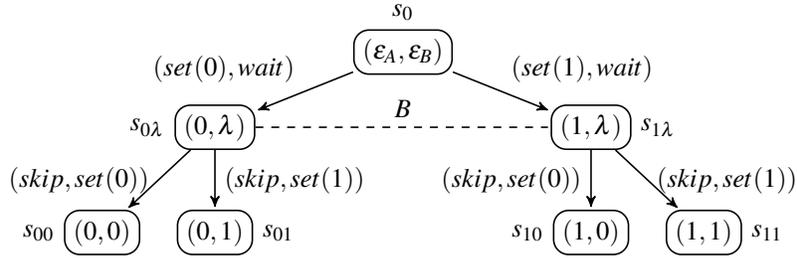

\textbf{Example.}
We introduce a turn-based \aqis\ with two agents, $A$
and $B$. First, $A$ secretly chooses between 0 and 1. Then, at the
successive stage, $B$ also chooses between 0 and 1. The game is won by agent
$A$ if the values provided by the two agents coincide, otherwise $B$
wins.
We formally describe this toy game starting with agents $A$ and $B$. Specifically, 
 $A$ is the tuple $\tup{L_A, Act_A, Pr_A}$, where
\myi $L_A = \{\epsilon_A, 0, 1 \}$;
\myii $Act_A = \{ set(0), set(1), skip \}$; and
\myiii $Pr_A(\epsilon_A) = \{ set(0), set(1) \}$ and $Pr_A(0) = Pr_A(1) =
  \{ skip \}$.
%
Further, agent $B$ is defined as the tuple $\tup{L_B, Act_B, Pr_B}$, where
$L_B = \{ \epsilon_B, \lambda, 0, 1 \}$;
$Act_B = \{ wait, set(0), set(1),  skip \}$;
$Pr_B(\epsilon_B) = \{ wait \}$, $Pr_B(\lambda) = \{ set(0), set(1) \}$
and $Pr_B(0) = Pr_B(1) = \{ skip \}$.
The intuitive meaning of local states, actions and protocol functions is clear.
Also, we consider the set $AP = \{ win_A, win_B \}$ of atomic propositions, which intuitively express that agent $A$ (resp.~$B$) has won the game. 
We now introduce the \aqis\ $\Q$, corresponding to our toy game, as
the tuple $\tup{Ag, s_0, \tau, \pi}$,
where
\myi $s_0 = (\epsilon_A,\epsilon_B)$;
\myii the transition function $\tau$ is given as follows for $i, j \in
  \{ 0,1 \}$:
\begin{itemize}
\item $\tau((\epsilon_A,\epsilon_B),(set(i),wait))= (i,\lambda)$
\item $\tau((i,\lambda),(skip,set(j)))= (i,j)$
\item $\tau((i,j),(skip,skip))= (\epsilon_A,\epsilon_B)$
\end{itemize}
and \myiii $\pi(win_A) = \{ (0,0), (1,1) \}$, $\pi(win_B) = \{ (1,0), (0,1) \}$.
Notice that we suppose that our toy game, represented in
Fig.~\ref{fig1}, is non-terminating. 

Now, we check whether the following ESL specifications hold in
the \aqis\ $\Q$.
\begin{eqnarray}
\Q & \models & \forall x_A \ X \ K_B \ \exists y_B \ X \ win_B \label{eq1}\\
\Q & \not \models & \forall x_A \ X \ \exists y_B \ K_B \ X \ win_B \label{eq2}\\
\Q & \models & \forall x_A \ X \ K_B \ K_A \ \exists y_B  \ X \ win_A \label{eq3}\\
\Q & \models & \forall x_A \ X \ K_B \  \exists  y_B \ K_A \ X \ win_A \label{eq4}
\end{eqnarray}

Intuitively, (\ref{eq1}) expresses the fact that at the beginning of
the game, independently from agent $A$'s move, at the next step agent
$B$ knows that there exists a move by which she can enforce her
victory. That is, if agent $A$ chose 0 (resp.~1), then $B$ can choose
1 (resp.~0). However, $B$ only knows that there exists a move, but she
is not able to point it out. In fact, (\ref{eq2}) does not hold, as
$B$ does not know which specific move $A$ chose, so she is not capable
of distinguishing states $s_{0\lambda}$ and $s_{1\lambda}$.  Moreover,
by (\ref{eq3}) $B$ knows that $A$ knows that there exists a move by
which $B$ can let $A$ win. Also, by (\ref{eq4}) this move is known to
$A$, as it is the $B$-move matching $A$'s move.

Indeed, in ESL it is possible to express the difference between {\em
de re} and {\em de dicto} knowledge of strategies. One of the first
contributions to tackle this issue formally is \cite{JamrogaH04}.
Formula~(\ref{eq1}) expresses agent $B$'s {\em de dicto} knowledge of
strategy $y_B$; while (\ref{eq2}) asserts {\em de re} knowledge of the
same strategy. Similarly, in (\ref{eq3}) agent $A$ has {\em de re}
knowledge of strategy $y_B$; while (\ref{eq4}) states that agent $A$
knows the same strategy {\em de dicto}.  The {\em de re}/{\em de
dicto} distinction is of utmost importance as, as shown above, having
a {\em de dicto} knowledge of a strategy does not guarantee that an
agent is actually capable of performing the associated sequence of
actions. Ideally, in order to have an effective strategy, agents must know it
 {\em de re}.


\section{Model Checking ESL} \label{MCesl}


In this section we consider the complexity of the model checking
problem for ESL.  In Section~\ref{hardness} and \ref{sec:ATA} we
provide the lower and upper bound respectively.  For reasons of space,
we do not provide full proofs, but only give the most important
partial results. We refer to \cite{Belardinelli14tech} for
detailed definitions and complete proofs.

 For an ESL formula $\phi$ we define $alt(\phi)$ as the
maximum number of alternations of quantifiers $\exists$ and $\forall$
in $\phi$.  Then, ESL[$k$-alt] is the set of ESL formulas $\phi$ with
$alt(\phi)$ equal to or less than $k$.


\subsection{Lower Bound} \label{hardness}

In this section we prove that model checking ESL formulas is
non-elementary-hard.  
Specifically, we show that for ESL formulas with maximum alternation
$k$ the model checking problem is $k$-EXPSPACE-hard. The proof
strategy is similar to \cite{MogaveroMV11}, namely,
we reduce the satisfiability problem for quantified propositional temporal logic (QPTL) to ESL model checking. 
However, the reduction applied is different, as ESL does not contain the bind operator used 
in \cite{MogaveroMV11}.

We first state that the satisfiability problem for QPTL sentences built
on a finite set $AP = \{ p_0, \ldots, p_n \}$ of atomic propositions
can be reduced to model checking ESL sentences on a \aqis\ $\Q$ of
fixed size on $|AP|$, albeit exponential.
%
\begin{lemma}[QPTL Reduction] \label{aux}
Let $AP = \{ p_0, \ldots, p_n \}$ be a finite set of atomic propositions.
There exists an \aqis\ $\Q$ on $AP$ s.t.~for every QPTL[$k$-alt]
sentence $\phi$ on $AP$, there exists an ESL[$k$-alt] sentence
$\overline{\phi}$ s.t. $\phi$ is satisfiable iff $\Q \models \overline{\phi}$.
\end{lemma}

By this result and the fact that the satisfiability problem for
QPTL[$k$-alt] is $k$-EXPSPACE-hard \cite{MogaveroMV11}, we can derive
the lower bound for model checking ESL[$k$-alt].
\begin{theorem}[Hardness] \label{hard}
The model checking problem for
ESL[$k$-alt] is $k$-EXPSPACE-hard.
\end{theorem}


In particular, it follows that ESL model checking is non-elementary-hard.

\subsection{Upper Bound} \label{sec:ATA}


In this section we extend to Epistemic Strategy Logic the model checking procedure for SL in \cite{MogaveroMV11}, which is based on
alternating tree automata (ATA) \cite{MullerS87}.
We state the following result, which extends Lemma 5.6
in \cite{MogaveroMV11}. 
\begin{lemma} \label{aux3}
   Let $\P$ be an \aqis\ and $\phi$ an ESL formula.  Then, there
   exists an alternating tree automaton $\A^{\phi}_{\P}$
s.t.~for every state $s \in S$
   and assignment $\chi$, we have that $(\P, s, \chi) \models \phi$
   iff the {\em assignment-state encoding} $\T^{\chi}_s$ belongs to the language  $\L(\A^{\phi}_{\P})$.
\end{lemma}

The following result corresponds to Theorem~5.4 in \cite{MogaveroMV11}.
\begin{theorem}[ATA Direction Projection]  \label{projection}
Let $\A^{\phi}_{\P}$
be the ATA
in Lemma~\ref{aux3},
and $s \in S$ a
   distinguished state.
Then, there
   exists a non-deterministic ATA $\N^{\phi}_{\P,s}$
s.t.~for all
   $Act_{\free(\phi)}$-labelled $\Delta$-tree $\T = \langle T,
   V \rangle$, we have that $\T \in \L(\N^{\phi}_{\P,s})$ iff
   $\T' \in \L(\A^{\phi}_{\P})$, where $\T'$ is the
   $(Act_{\free(\phi)} \times S)$-labelled $\Delta$-tree $\langle T,
   V' \rangle$ 
s.t.~$V'(x) = (V(x), last(\kappa_{s \cdot x}))$.
\end{theorem}

Then, by using Lemma~\ref{aux3} and Theorem~\ref{projection} we can state the following result.  
\begin{theorem}
Let $\P$ be an \aqis, $s$ a state in $\P$, $\chi$ an assignment, and
$\phi$ an ESL formula. The non-deterministic ATA $\N^{\phi}_{\P,s}$ in
Theorem~\ref{projection} is such that $(\P, s , \chi) \models \phi$ iff
$\L(\N^{\phi}_{\P,s}) \neq \emptyset$.
\end{theorem}

We can finally state the following extension to Theorem 5.8
in \cite{MogaveroMV11}, which follows from the fact that the
non-emptyness problem for alternating tree automata is non-elementary
in the size of the formula.
\begin{theorem}[Completeness]  \label{comple}
The model checking problem for ESL is PTIME-complete w.r.t.~the size of
the model and NON-ELEMENTARYTIME w.r.t.~the size of the formula.
\end{theorem}


We remark that Theorem~\ref{comple} can be used to show
that the model checking problem for the
nested-goal fragment ESL[NG]
 is PTIME-complete w.r.t.~the size of
the model and ($k+1$)-EXPTIME w.r.t.~the maximum  alternation $k$ of a formula. 
We conclude that the complexity of model checking ESL is not worse than the corresponding problem for the Strategy Logic in \cite{MogaveroMV11}.

\section{Conclusions} \label{conc}


In this paper we introduced Epistemic Strategy Logic, an extension of
Strategy Logic \cite{MogaveroMV11} with modalities for individual
knowledge. We provided this specification language with a semantics in
terms of epistemic concurrent game models (\aqis), and analysed the
corresponding model checking problem.  A number of developments for
the proposed framework are possible. Firstly, the model checking
problem for the nested-goal, boolean-goal, and one-goal fragment of SL
has lower complexity. It is likely that similar results hold also for
the corresponding fragments of ESL. Secondly, we can extend ESL with
modalities for group knowledge, such as common and distributed
knowledge. Thirdly, we can consider various assumptions on \aqis, for
instance perfect recall, no learning, and synchronicity. The latter
two extensions, while enhancing the expressive power of the logic, are
also likely to increase the complexity of the model checking and
satisfiability problems.
  


 
\bibliographystyle{eptcs}
\bibliography{sr2014}








\end{document}